\documentclass[journal=jaaucr,manuscript=article]{achemso}

\usepackage{chemformula} 
\usepackage[T1]{fontenc} 
\usepackage{braket}
\usepackage{mathtools}
\usepackage{amssymb}
\usepackage{threeparttable}
\usepackage{bbm}
\usepackage{leftindex}
\usepackage{listings}  
\usepackage{multirow}
\DeclareMathOperator{\Tr}{Tr}

\newcommand{\Tthree}[4][M]{{#1}_{#2,#3}^{#4}}
\newcommand{\Tfour}[5][M]{{#1}_{#2,#3}^{#4,#5}}

\author{Yifan Cheng}
\affiliation[Nanjing University]
{School of Chemistry and Chemical Engineering,
Nanjing University, Nanjing, Jiangsu 210023, China}
\author{Zhaoxuan Xie}
\affiliation[LMU]
{Department of Physics and Arnold Sommerfeld Center for Theoretical Physics (ASC), Ludwig-Maximilians-Universität München, Theresienstr. 37, München D-80333, Germany}
\author{Xiaoyu Xie}
\email{xiaoyuxie@sdu.edu.cn}
\author{Haibo Ma}
\email{haibo.ma@sdu.edu.cn}
\affiliation[Shandong University]
{Qingdao Institute for Theoretical and Computational Sciences, School of Chemistry and Chemical Engineering, Shandong University, Qingdao, Shandong 266237, China}

\title[BIPS-DMRG]
  {Efficient simulation of inhomogeneously correlated systems using block interaction product states}


\begin{document}

\begin{tocentry}
\includegraphics[width=0.95\textwidth]{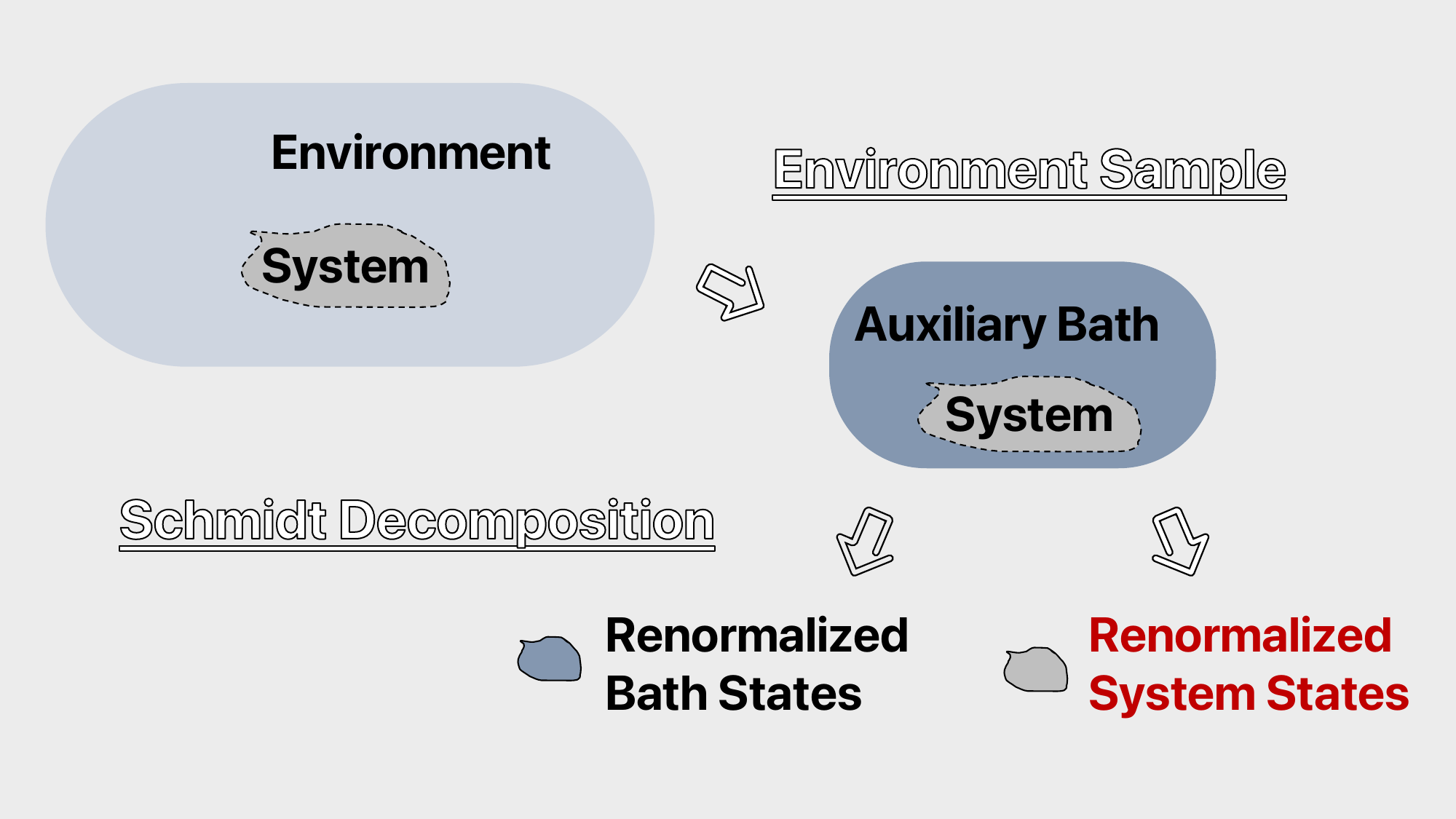}
\end{tocentry}

\begin{abstract}
The strength of the density matrix renormalization group (DMRG) in handling strongly correlated systems lies in its unbiased and simultaneous treatment of identical sites that are both energetically degenerate and spatially similar, as typically encountered in physical models. However, this very feature becomes a drawback when DMRG is applied to quantum chemistry calculations for large, realistic correlated systems. This is because entangled orbitals often span broad ranges in both energy and space, with their interactions being notably inhomogeneous. In this study, we suggest addressing the strong intra-fragment correlations and weak inter-fragment correlations separately, utilizing a large-scale multi-configurational calculation framework grounded in the block interaction product state (BIPS) formulation. The strong intra-fragment correlation can be encapsulated in several electronic states located on fragments, which are obtained by considering the entanglement between fragments and their environments. Moreover, we incorporate non-Abelian spin-SU(2) symmetry in our work to target the desired states we interested with well-defined particle number and spin, providing deeper insights into the corresponding chemical processes. The described method has been examined in various chemical systems and demonstrates high efficiency in addressing the inhomogeneous effects in strong correlation quantum chemistry.
\end{abstract}

\section{Introduction}
Thanks to the high compression of the matrix product state (MPS) representation of wave function and the efficient site-by-site iterative sweeping optimization algorithm, the density matrix renormalization group (DMRG) method \cite{whiteDensityMatrixFormulation1992,whiteDensitymatrixAlgorithmsQuantum1993,schollwoeckDensitymatrixRenormalizationGroup2005,schollwaDensitymatrixRenormalizationGroup2011,xiangDensitymatrixRenormalizationgroupMethod1996a,verstraeteDensityMatrixRenormalization2004} has established itself as a powerful computational tool for accurately simulating one-dimensional (1D) strongly correlated systems with only local interactions and nearly identical sites. While such situations are typically encountered in physical models such as the Heisenberg \cite{depenbrockNatureSpinLiquidGround2012a,zhuSpinLiquidPhase2015} or Hubbard \cite{zhengStripeOrderUnderdoped2017,arovasHubbardModel2022} Hamiltonians, large and realistic correlated systems often exhibit inhomogeneous electron correlations in the realm of quantum chemistry. For instance, in conjugated aggregates and polymers, the correlation within a single monomer fragment is typically more pronounced than the inter-fragment correlation between different monomer fragments. Since the $ab$ $initio$ quantum chemistry Hamiltonian encompasses numerous long-range interactions and four-operator terms, quantum chemistry DMRG (QC-DMRG) calculations \cite{whiteInitioQuantumChemistry1999,chanMatrixProductOperators2016,daulFullCIQuantumChemistry2000,ma2022density,chan2011density} scale steeply with system size and necessitate a substantial number of truncated renormalized bases. Given their vast size, many chemical systems with inhomogeneous correlations, such as polymers, molecular aggregates, and polynuclear transition metal complexes, are far beyond the current capabilities of QC-DMRG. Hence, there is a pressing need to develop novel efficient methods tailored to these inhomogeneously correlated systems.

To address correlations of varying strengths, one can resort to designing specialized tensor networks (TNs) that assign sites with different entangling environments to distinct node levels. One example is the comb tensor network (CTN) proposed by Chepiga and White in 2019 \cite{chepigaCombTensorNetworks2019,chepigaCriticalPropertiesComb2020}, which features a 1D backbone with finite 1D teeth originating from it, representing different magnitudes of correlation separately. However, its non-1D TN form brings high computational costs and complicates the construction of the quantum chemistry Hamiltonian. As a result, even though the expressibility of CTN state (CTNS) has been investigated for the P-cluster and the FeMoco of nitrogenase by Li\cite{li2021expressibility}, its efficient implementation for quantum chemical systems has not been reported so far. Besides, some works on tree tensor network states (TTNS) for quantum chemistry have also been proposed, which can be referenced in these pieces of literature. \cite{orus2019tensor,nakatani2013efficient,szalay2015tensor}.


Other approaches to address the inhomogeneous interaction issue have mainly focused on the coarse-graining idea, which groups the neighboring strongly correlated sites into one single coarse grained large site, sometimes called as ”cluster”, to pre-package strong intra-fragment correlation and simplify the simulation of the whole system. Methods like the cluster mean field (cMF) \cite{jimenez-hoyosClusterbasedMeanfieldPerturbative2015, papastathopoulos2024linear} or localized active space self-consistent field (LASSCF) \cite{hermesMulticonfigurationalSelfConsistentField2019,hermes2020variational} describe the intra-fragment correlation at the full configurational interaction (FCI) level while treating weaker inter-fragment correlations with mean-field theory, and can be viewed as a cluster version of Hartree-Fock (HF) theory. Consequently, there remains a need to incorporate the neglected inter-fragment correlations, akin to how electron correlation is treated beyond standard HF. Many approaches have been proposed, including active-space decomposition (ASD) \cite{parker2013communication,kim2015orbital}, rank-one basis strategy \cite{nishio2019rank}, cMF-perturbation theory (PT) \cite{jimenez-hoyosClusterbasedMeanfieldPerturbative2015}/coupled-electron pair approximation (CEPA) \cite{abraham2022coupled}/coupled cluster (CC) \cite{papastathopoulos2022coupled}, cluster many-body expansion (cMBE) \cite{abraham2021cluster}, generalized cMF (GcMF) \cite{papastathopoulos2023symmetry}, LAS state interaction (LASSI) \cite{agarawal2024automatic,pandharkar2022localized}, and tensor product state configuration interaction (TPSCI) \cite{abraham2020selected,braunscheidel2023generalization}. These methods correspond to PT2, CEPA, CC, many-body expansion (MBE), generalized HF (GHF), and selected configuration interaction (CI) methods that aim to account for electron correlation beyond HF theory. Notably, the generalized valence bond block correlated PT/CC (GVB-BCPT/CC) proposed by Li et al. \cite{xu2013block,zou2022efficient,ren2024block}, which involves fragments consisting of just two orbitals, shares similarities with these methods due to the localized nature of GVB orbitals.

The application of these standard electron correlation methods to address inter-fragment correlations is not trivial. The large local Hilbert spaces spanned by the configurational bases of coarse-grained sites make direct implementation very time-consuming. Tucker decomposition can be utilized within variational optimization to compress the Hilbert space \cite{zhangDensityMatrixApproach1998a,mayhall2017using}, but it still requires exploration of the entire local Hilbert space during the optimization. A more common approach is to keep only a few energetically low-lying adiabatic states to represent the local Hilbert space. However, the local adiabatic states derived from an isolated fragment's calculation cannot adequately capture the realistic inter-fragment interactions, just like the well-known breakdown of numerical renormalization group (NRG) methods for correlated systems \cite{wilson1975renormalization}. Moreover, for different chemical problems, the required local adiabatic states, which vary in electron number and spin, will differ. Their proportions and quantities need to be manually determined or set according to several predefined rules based on chemical insights\cite{auto_las}. Therefore, it is crucial to develop a method that can automatically generate optimal local state bases that adequately account for environmental effects. Here, "environmental effects" refer to the influences of other fragments on the target fragment and are distinguished from the term "environmental effect" used in the context of multi-scale approaches\cite{gordon2001effective}.

In this work, we introduce a novel coarse-grained method designed to efficiently simulate inhomogeneously correlated systems using a block interaction product state (BIPS) formulation \cite{wangLowScalingExcitedState2022}. For each fragment, a compact set of renormalized local states is optimally constructed while accounting for environmental effects. This is achieved by performing one-shot density matrix embedding theory (DMET) \cite{kniziaDensityMatrixEmbedding2012,knizia2013density} to compress the environment space into an auxiliary bath space, which has the same size as the fragment. The set of renormalized local states can be automatically selected by performing the Schmidt decomposition of the wave function in a model space composed of the fragment and its bath space, with a defined maximum number of total states or a threshold for discarded singular values.

For treating inter-fragment interactions, we utilize the cluster DMRG (cDMRG) method \cite{whiteDensityMatrixFormulation1992,largesite} with a small bond dimension. This procedure can be seamlessly integrated with BIPS, offering an accurate and computationally efficient way to describe inhomogeneously correlated systems. Compared to the previously mentioned treatments of inter-fragment correlation, cDMRG inherits the traits of standard DMRG, sampling the entire Hilbert space according to entanglement and efficiently addressing weak but homogeneous inter-fragment correlations. Additionally, the non-Abelian spin-SU(2) symmetry can be easily incorporated into the cDMRG method, resulting in a spin-adapted version of this approach. Unlike previous coarse-grained methods that use spin-raising and lowering operators to reduce spin contamination \cite{parker2014quasi,parker2014model,braunscheidel2023generalization}, our method not only produces diabatic states instead of adiabatic states but also ensures that these states are spin-adapted for the first time. Combined with the sampling algorithm \cite{leeExternallyCorrectedCCSD2021} adopted from standard DMRG, the defined electronic states in BIPS-DMRG are more physically meaningful and provide better insight into the corresponding physical and chemical processes.

\section{Theory and Methodology}
\subsection{DMRG Basics}
Considering a system composed of $k$ sites, each of which have $d$ local states. The wave function can be expressed in terms of MPS structures with different canonical forms, considering the gauge freedom of TNs. Assuming the orthogonality center site $p$ divide the chain into left system and right environment blocks, then the wave function can be written in the mixed canonical form as:\cite{whiteDensityMatrixFormulation1992,whiteDensitymatrixAlgorithmsQuantum1993,schollwoeckDensitymatrixRenormalizationGroup2005,schollwaDensitymatrixRenormalizationGroup2011,xiangDensitymatrixRenormalizationgroupMethod1996a,verstraeteDensityMatrixRenormalization2004}
\begin{equation}
\begin{aligned}
    \ket{\Psi} &= 
    \sum_{\{\sigma_{i}\}}\sum_{\{a_i=1\}}^{\{m_i\}} 
    \Tthree[U]{1}{a_1}{\sigma_{1}}\cdots \Tthree[M]{a_{p-1}}{a_{p}}{\sigma_{p}}
    \cdots\Tthree[V]{k-1}{1}{\sigma_{k}} \ket{\sigma_{1}\cdots\sigma_{p}\cdots\sigma_{k}}\\
    &=\sum_{\sigma_{p}}\sum_{a_{p-1},a_{p}} (\Tthree{a_{p-1}}{a_{p}}{\sigma_{p}} \ket{L_{a_{p-1}}} \otimes\ket{\sigma_{p}})\otimes \ket{R_{a_{p}}} \\
    &=\sum_{a_{p}=1}^{m_{p}} S_{a_{p}} \ket{L_{a_{p}}} \otimes \ket{R_{a_{p}}},
\label{eq:mixed-canon}
\end{aligned}
\end{equation}
in which $U/V/\Tthree{a_{i-1}}{a_i}{\sigma_{i}}$ represents an element of the left-normalized/right-normalized/general local rank-3 tensor $\bf{U/V/M}$, which has two auxiliary bonds $ a_{i-1}$, $ a_i$ with dimensions ${m}_{i-1}$, $m_i$ and one physical bond ${\sigma}_i$ with dimension $d$. The $\bf{U}$ and $\bf{V}$ originate from the singular value decomposition (SVD) procedure and the sets $\{\ket{L_{a_{p}}}\}$ and $\{\ket{R_{a_{p}}}\}$ denote renormalized states of system and environment blocks, respectively. Since $\{\ket{L_{a_{p}}}\}$ are eigenvectors of the reduced density matrix (RDM) for system block, meaning that the RDM is diagonal when represented in such bases:
\begin{equation}
\begin{aligned}
    \rho_{\text{sys}} = \Tr_{\text{env}} (|\Psi\rangle \langle \Psi|)
       = \sum_{a_{p}=1}^{m_{p}}
        \langle R_{a_{p}} |\Psi\rangle \langle \Psi|  R_{a_{p}} \rangle
        =\sum_{a_{p}=1}^{m_{p}}
        S^2_{a_{p}} \ket{L_{a_{p}}}  \bra{L_{a_{p}}}.
\label{eq:RDM}
\end{aligned}
\end{equation}
Therefore, these renormalized states can be seemed as the most compact many-particle bases to represent the Hilbert space of system block, which is also the key ingredient behind the success of DMRG.

Similarly, one can decompose the global operator $\widehat{O}$ to the MPO tensors,
\begin{equation}
    \label{eq:general_mpo}
    \widehat{O} = \sum_{\{{\sigma}_i {\sigma}^{*}_i\}}\sum_{\{b_i\}}
    {\Tfour[W]{1}{b_{1}}{\sigma^{*}_1}{\sigma_1}
    \Tfour[W]{b_{1}}{b_{2}}{\sigma^{*}_2}{\sigma_2}
    \cdots
    \Tfour[W]{b_{k-1}}{1}{\sigma^{*}_k}{\sigma_k} \vert{\sigma}^{*}_1\cdots{\sigma}^{*}_k\rangle\langle{\sigma}_1\cdots{\sigma}_k\vert}
\end{equation}
in which $\Tfour[W]{b_{i-1}}{b_i}{\sigma^{*}_i}{\sigma_i}$ represents an element of the local rank-4 tensor ${W}[i]$, which has two link bonds $ b_{i-1}$, $ b_i$ with dimensions ${D}_{i-1}$, $D_i$ and two physical bonds ${\sigma}_i$, ${\sigma}^{*}_i$. Using $\partial E/\partial M_{i}=0$, the final working equation for optimizing $M_{i}$ can be easily derived as:
\begin{align}
     \hat{H}_{i}^{\mathrm{eff}}M_{i} = EM_{i}.
    \label{eq:diag}
\end{align}
Here $\hat{H}_{i}^{\mathrm{eff}}\in\mathbb{C}^{dm^{2}\times dm^{2}}$ is single-site effective Hamiltonian acting on the projected space $\{|L_{a_{i-1}}\sigma_{i}R_{a_{i}}\rangle\}$. To explore a larger variational space ($\{|L_{a_{i-1}}\sigma_{i}\sigma_{i+1}R_{a_{i+1}}\rangle\}$ with dimension $d^2m^{2}$) to avoid possible trapping in local minima, the two-site algorithm can be employed by introducing the two-site tensor $T_{i,i+1}$:
\begin{align}
   T^{\sigma_{i},\sigma_{i+1}}_{a_{i-1},a_{i+1}} = \sum_{a_{i}}M^{\sigma_{i}}_{a_{i-1},a_{i}}M^{\sigma_{i+1}}_{a_{i},a_{i+1}}.
   \label{eq:2site}
\end{align}

It is evident that, the computational accuracy and cost of the DMRG method can be controlled by adjusting the value of $m$. Although it is principally possible to simulate inhomogeneous correlation using DMRG with a variable bond dimension--for example, by using dynamical block state selection (DBSS) \cite{legezaControllingAccuracyDensitymatrix2003} to determine auxiliary states based on singular values--the efficient application of DMRG to chemical systems is still limited by high computational costs. The factors of $k^4/k^3$ in a long chain with $k$ sites and $m^2/m^3$ for large $m$ values collectively lead to prohibitive computational expenses, $O(k^4m^2) + O(k^3m^3)$, in QC-DMRG for large chemical systems.\cite{whiteInitioQuantumChemistry1999,chanMatrixProductOperators2016,daulFullCIQuantumChemistry2000,ma2022density,chan2011density}

\subsection{BIPS}
Considering the expensive computational cost of DMRG for large chemical systems, we suggest addressing the inhomogeneous electron correlation (the strong intra-fragment correlation and weak inter-fragment correlation) separately. By fusing the orbitals that belong to the same strongly correlated regions together, the overall wave function can be formulated as:
\begin{equation}
\begin{aligned}
    |\Psi\rangle &= \sum_{\alpha,\beta,\cdots,\omega}c_{{\alpha_1},\beta_2,\cdots,\omega_N}|\alpha_1\beta_2\cdots\omega_N\rangle\\
    &= \sum_{\alpha,\beta,\cdots,\omega}\sum_{\{\Tilde{a}_I=1\}}^{\{\Tilde{m}_I\}} {\Tthree{1}{\Tilde{a}_{1}}{\alpha_{1}}\Tthree{\Tilde{a}_{1}}{\Tilde{a}_{2}}{\beta_{2}}\cdots \Tthree{\Tilde{a}_{N-1}}{1}{\omega_{N}}|{\alpha_1\beta_2\cdots\omega_N}}\rangle.
    \label{eq:BIPS_form}
\end{aligned}
\end{equation}
In this equation, $N$ is the number of fused regions (blocks/clusters) in the system, and Greek letters label the states of a particular region. The indices $\{\Tilde{m}_I\}$ and $\{\Tilde{a}_I\}$ are used to distinguish them from $\{m_i\}$ and $\{a_i\}$ in the standard MPS. The upper limit of $\alpha, \beta, \dots, \omega$ is determined by the dimension of the local basis set. At this maximal setting, this generalized formula is quite flexible in choosing local states, as transformations within the same region do not affect the energy. However, if the number of local states is reduced, the choice of local states will introduce different errors when approximating the true wave function. For example, if the local Hilbert space is spanned by only one state, the formula yields the cMF/LAS wave function. The optimized local state obtained from the cMF/LASSCF calculation would then produce the smallest error in representing the true wave function in this context.




\begin{figure}[htbp]
\includegraphics[width=\textwidth]{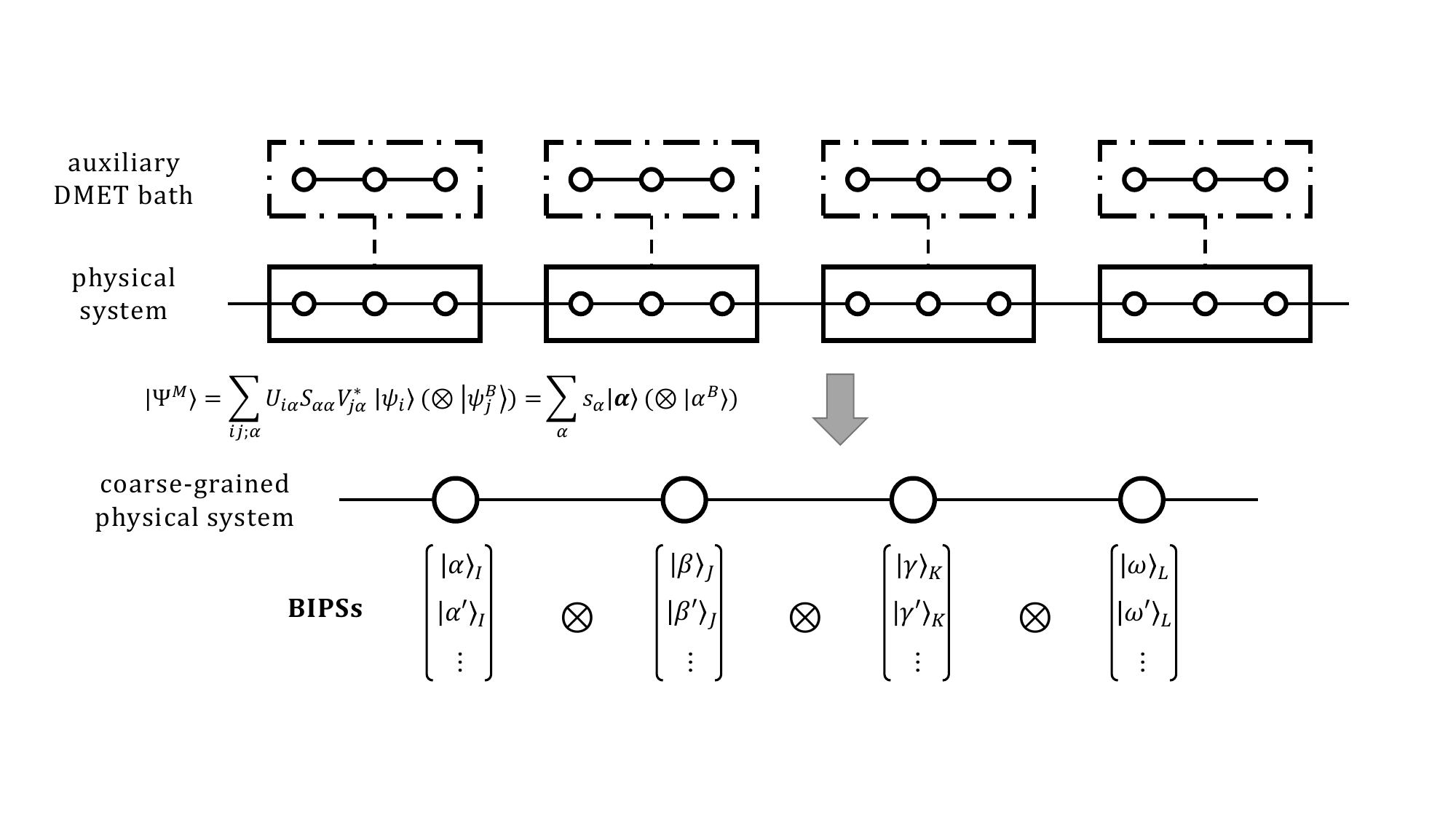}
\caption{\label{fig:BIPS}Algorithms for constructing BIPS utilize the auxiliary DMET bath environment. In the equation, the superscripts $M$ and $B$ denote the model space and bath space, respectively. The bath states are used solely to assist in constructing local states and do not explicitly appear in the BIPS wave function.}
\end{figure}

Principally, the form in eq~\ref{eq:BIPS_form} holds true for any system when enough local basis states are used, regardless of the specific choices of these states. However, we find that using only a few appropriately chosen states, which take into account system-environment effects, can already achieve chemical accuracy for numerous molecular systems of interest. These selected states encapsulate strong correlations within the system and yield a compact representation of the local Hilbert space. The product of these local states are herein named as block interaction product states (BIPSs).

Our algorithm for obtaining BIPSs begins with localized and orthogonal molecular orbitals $\{\phi_i^{LO}\}$ (Figure~\ref{fig:BIPS}). By partitioning system into $N$ block fragments, these orbitals can be assigned to each block according to their dominant atomic orbital contributions. For each block fragment, a set of bath orbitals is generated to simulate the environmental influences on the construction of local states. This set should be small to avoid costly calculations of the model system, which includes the fragment of interest and its defined bath orbitals. In this case, we utilize one-shot DMET \cite{kniziaDensityMatrixEmbedding2012,knizia2013density}, performing orbital rotation only once without further self-consistent optimization, to generate a bath orbital space that has only the same size as the local fragment orbital space for each individual fragment. In practice, the 1-particle reduced density matrix (1-RDM) in local fragment orbital $\{\phi_i^{LO}\}$ bases can be obtained from a low-level solver, usually the HF, as:
\begin{equation}
\begin{aligned}
        P_{ij}=\langle \Phi |\hat{a}_j^\dagger \hat{a}_i|\Phi\rangle=\sum_\mu^{N_{occ}} C_{i\mu}C_{j\mu}^\dagger
\end{aligned}
\end{equation}
 where $i$ and $j$ denote the local fragment orbitals, and $\mu$ denotes the occupied canonical orbital. Here, $\Phi$ represents the Slater determinant. Since the 1-RDM is non-diagonal in the local orbital bases, we can focus on the $k_I \times (k - k_I)$ subblock corresponding to fragment $I$, where $i$ is in fragment $I$ and $j$ in other fragments. Performing SVD of this subblock yields $k_I$ bath orbitals and $k - 2k_I$ unentangled environment orbitals. To construct optimal local states, the model system of $2k_I$ fragment and bath orbitals needs to be treated at a high-correlated level, such as DMRG, TN, or FCI, to obtain the wave function $|\Psi^{M}\rangle_I$. Subsequently, the fragment's RDM can be constructed from $|\Psi^{M}\rangle_I$ by tracing out the bath degrees of freedom, and then diagonalized to obtain the renormalized local state bases with large eigenvalues:
 \begin{equation}
 \begin{aligned}
     \rho_{I} &= \sum_{j,j'}
        \leftindex_{I}{\langle \leftindex^{B} \psi_j |} {\Psi^{M}\rangle}_I \leftindex_{I}{\langle \leftindex^{M} \Psi}|  {\psi_{j'}^{B} \rangle}_I\\
        &=\sum_{i,i';\alpha} U_{i\alpha} \lambda_{\alpha} U_{\alpha i'}{|\psi_i\rangle}_I \leftindex_I{\langle \psi_{i'} |}
        = \sum_{\alpha}
        \lambda_{\alpha} \ket{\alpha}_I  \leftindex_I {\bra{\alpha}}.
 \end{aligned}
 \end{equation}
Here, the superscripts $M$ and $B$ denote the model space and bath space, $\lambda$ represents the eigenvalue of $\rho_{I}$, and $U$ is the orthogonal transformation matrix that diagonalizes $\rho_{I}$ and can also be obtained from the Schmidt decomposition of $|\Psi^{M}\rangle_I$ (as shown in Figure~\ref{fig:BIPS}). Note that two types of Schmidt decomposition/SVD procedures are performed in the framework of BIPS-DMRG. The first is applied to the HF wave function using local orbital bases to define bath orbitals, while the second is applied to the wave function within the model system to define renormalized local states.
 
 Contrary to selecting local adiabatic bases based on chemical empirical intuition, these renormalized local states are determined automatically and include environmental effects. It is noteworthy that these eigenstates preserve symmetries of the group during the diagonalization of RDM and can be labelled with quantum number information. For instance, preserved states can be categorized into three groups according to charge symmetry: neutral states (neu), positively charged states (pos), and negatively charged states (neg). Therefore, the charge transfer or doubly excited states can be easily investigated by constructing particular BIPSs as the direct products of these local states. 
 
 

\subsection{BIPS-DMRG}
\begin{figure}
     \centering
     \includegraphics[width=\textwidth]{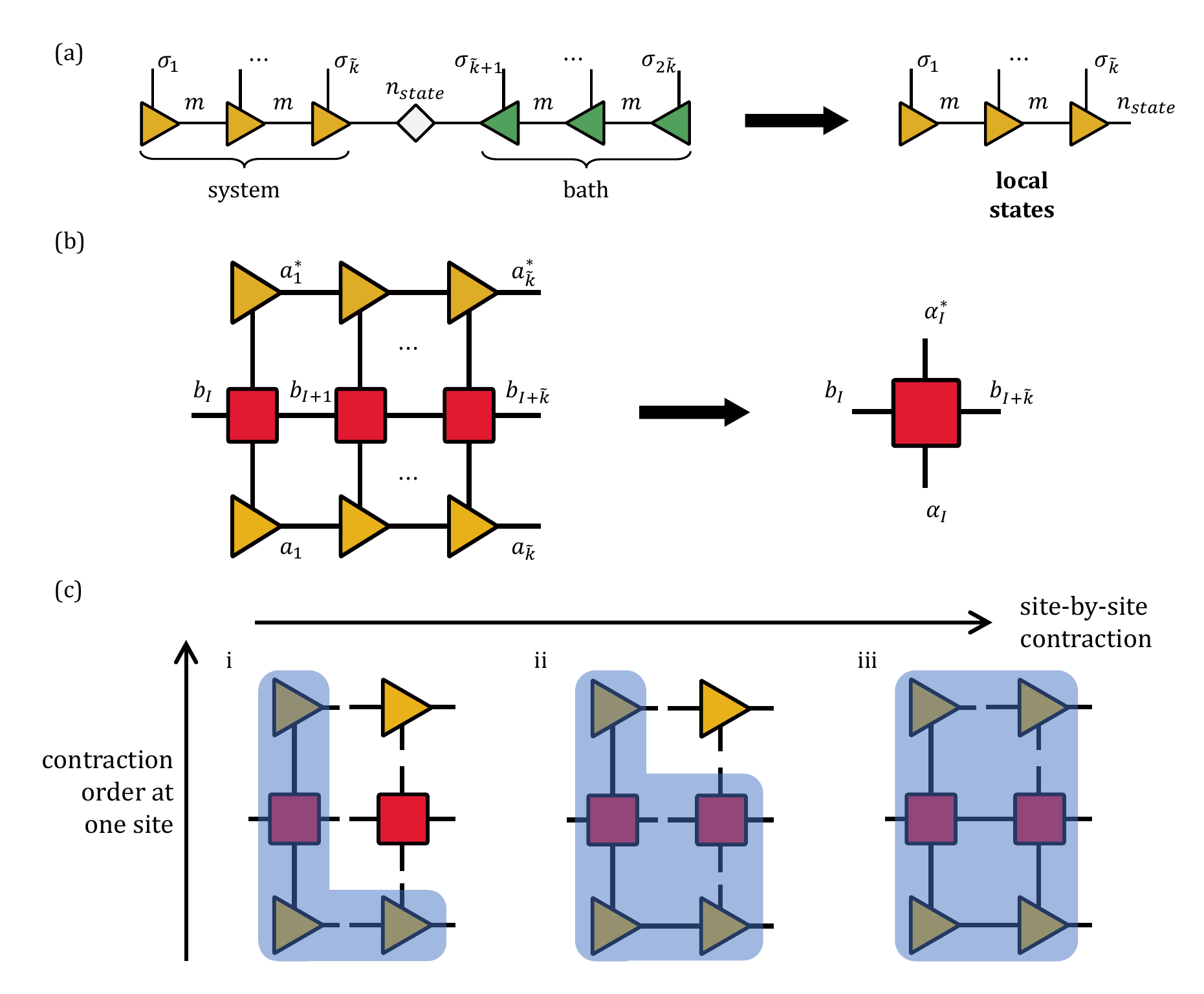}
     \caption{Schematic diagram of: (a) the bond-canonical formalism, a special case of the mixed canonical MPS formalism with the orthogonal center on the link bond, used to obtain local states in the MPS formalism; (b) the clustering procedure of MPO in BIPS-DMRG for fragment \( I \), comprising \(\tilde{k}\) orbitals (where \(\alpha_I\) is used as an alternative notation for \( a_{\tilde{k}} \) to represent the local state index in the cMPO); and (c) the optimal contraction order for cMPO. Red squares denote MPO, yellow and green triangles indicate left- and right-normalized MPS, respectively, and the gray diamond represents the singular value matrix. The regions highlighted in light blue indicate where tensor network contraction occurs. }
    \label{fig:clst_MPO}
 \end{figure}

As mentioned in the last section, local bases can be easily generated by performing Schmidt decomposition of the wave function within the model space. However, the obtained bases will be linear combinations of all possible configurations in the local basis set, which complicates the calculation of Hamiltonian elements between BIPSs. Therefore, we choose DMRG as the primary method for high-level calculations and generate the local bases using the highly compressed MPS structure (shown in Figure~\ref{fig:clst_MPO}.a). This can be easily achieved within DMRG by setting the bond dimension between the system and bath to the number of desired local states and placing the orthogonal center on the link bond connecting them. By representing these local states as MPSs, we can directly derive the required operator quantities localized in each cluster through tensor contraction and represent the total Hamiltonian as a cluster MPO (cMPO), where the physical bonds correspond to the selected local states. As a result, the calculation of Hamiltonian elements simplifies to tensor contractions, rather than computing numerous Hamiltonians between configurations, leading to a reduction in computational cost.


The construction procedure of the cMPO is similar to the propagation procedure in standard DMRG, as shown in Figure~\ref{fig:clst_MPO}.b, which also follows a site-by-site contraction approach. For a particular fragment \( I \), the procedure starts with the transformation of the first primitive MPO and then sequentially contracts with the MPSs and MPOs at the following sites. Figure~\ref{fig:clst_MPO}.c presents the optimal contraction order to minimize overhead. Note that in procedures i and iii of Figure~\ref{fig:clst_MPO}.c, for an arbitrary site \( p \) in fragment \( I \), the most time-consuming part is the transformation of one- and two-operator terms multiplied by the two-electron integrals\cite{keller2016efficient}, such as the \( v_{pqrs}a_p^+ \) quantities. Here, \( v_{pqrs} \) represents the two-electron integrals in the Hamiltonian, where indices \( q, r, s \) correspond to orbitals not belonging to fragment \( I \), and \( p \) denotes the current site. Since there are \( O(k^3) \) terms of such quantities at site \( p \), the transformation of these quantities would take \( O(k^3 m^3) \). Given that this transformation is performed at every site, the total computational cost would be \( O(k^4 m^3) \), which exceeds the computational cost of standard DMRG. Although this procedure only needs to be performed once and can be efficiently parallelized over different fragments, it will still become a bottleneck when the number of sites becomes too large. Therefore, a technique is needed to avoid this problem. One can first propagate the operators without the electron integrals, i.e., \(a_p^+\), and then multiply the integrals with them after propagation. The reason for this is that in BIPS-DMRG, the bond dimension used to represent the local states will be much larger than the number of local states. Consequently, after the transformation in each fragment, the dimension of the operator matrix in the cMPO will be quite small, which reduces the total computational cost of constructing the cMPO from \(O(k^4m^3)\) to \(O((k\tilde{k}^2 + k^2)m^3) + O(k^4n_{state}^2)\). Here, \(n_{state}\) is the number of local states, \(\tilde{k} \coloneqq k/N\) represents the number of sites within one fragment, and \(N\) is the number of fragments. The \(k\tilde{k}^2m^3\) term originates from the transformation of two-operator terms, such as \(a_o^+a_p\); the \(k^2m^3\) term comes from the three-body complementary operator terms\cite{chanMatrixProductOperators2016}, such as \(v_{nopq}a_n^+a_o^+a_p\); and the \(k^4n_{state}^2\) term results from the final multiplication with integrals to obtain such \(v_{pqrs}a_p^+\) quantities (where \(n\), \(o\), and \(p\) belong to fragment \(I\)).

 When considering the non-Abelian spin-$\mathrm{SU(2)}$ symmetry, the tensors that appear during propagation will consist of reduced matrix elements. During tensor contraction, the coupling coefficients must be multiplied, and as a result, the formulation can be expressed as follows:
\begin{equation}
\begin{aligned}
      (\Tfour[W]{b_{I}}{b_{I+i}}{a^*_i}{a_i})^{[k]}
      &=
      \sum_{\substack{\sigma^*_i,\sigma_i,\\a^*_{i-1},a_{i-1},\\b_{I+i-1}}}
      (-1) ^{S_{b_I} + S_{b_{I+i}} + k_1 + k_2 }
      {\sqrt{(2S_{b_{I+i-1}}+1)(2k+1)}} 
    \times  \begin{Bmatrix}
        S_{b_I} & k_1 & S_{b_{I+i-1}}\\
        k_2 & S_{b_{I+i}} & k
    \end{Bmatrix}\\
    &\times
  \begin{bmatrix}
    S_{a_{i-1}} & S_{\sigma_i} & S_{a_i} \\
    k_1 & k_2 & k \\
    S_{a^*_{i-1}} & S_{\sigma^*_i} & S_{a^*_i}\\
  \end{bmatrix}
    (\Tfour[W]{b_{I}}{b_{I+i-1}}{a^*_{i-1}}{a_{i-1}})^{[k_1]}
      (\Tthree[U]{a_{i-1}}{a_i}{\sigma_{i}})^*
      (\Tfour[W]{b_{I+i-1}}{b_{I+i}}{\sigma_{i}*}{\sigma_i})^{[k_2]}
      \Tthree[U]{a_{i-1}}{a_i}{\sigma_{i}}
      \label{eq:clust_mpo_prop_su2}
\end{aligned}
\end{equation}
Here, $ (\Tfour[W]{b_{I}}{b_{I+i}}{a^*_i}{a_i})^{[k]}$ is an element of the \(i\)-th primitive MPO in fragment $I$ and $\Tthree[U]{a_{i-1}}{a_i}{\sigma_{i}}$ is an element of the \(i\)-th primitive MPS representing local bases of fragment $I$. The $S_x$ represents the spin quantum number of index \(x\), \(k\) denotes the spin quantum number of $(\hat{W})^{[k]}$ itself, the quantity in curly brackets is a Wigner-6j symbol, and the square bracketed term is the product of a Wigner-9j symbol and a normalization factor \cite{keller2016spin}.

\begin{figure}
    \centering
    \includegraphics[width=\textwidth]{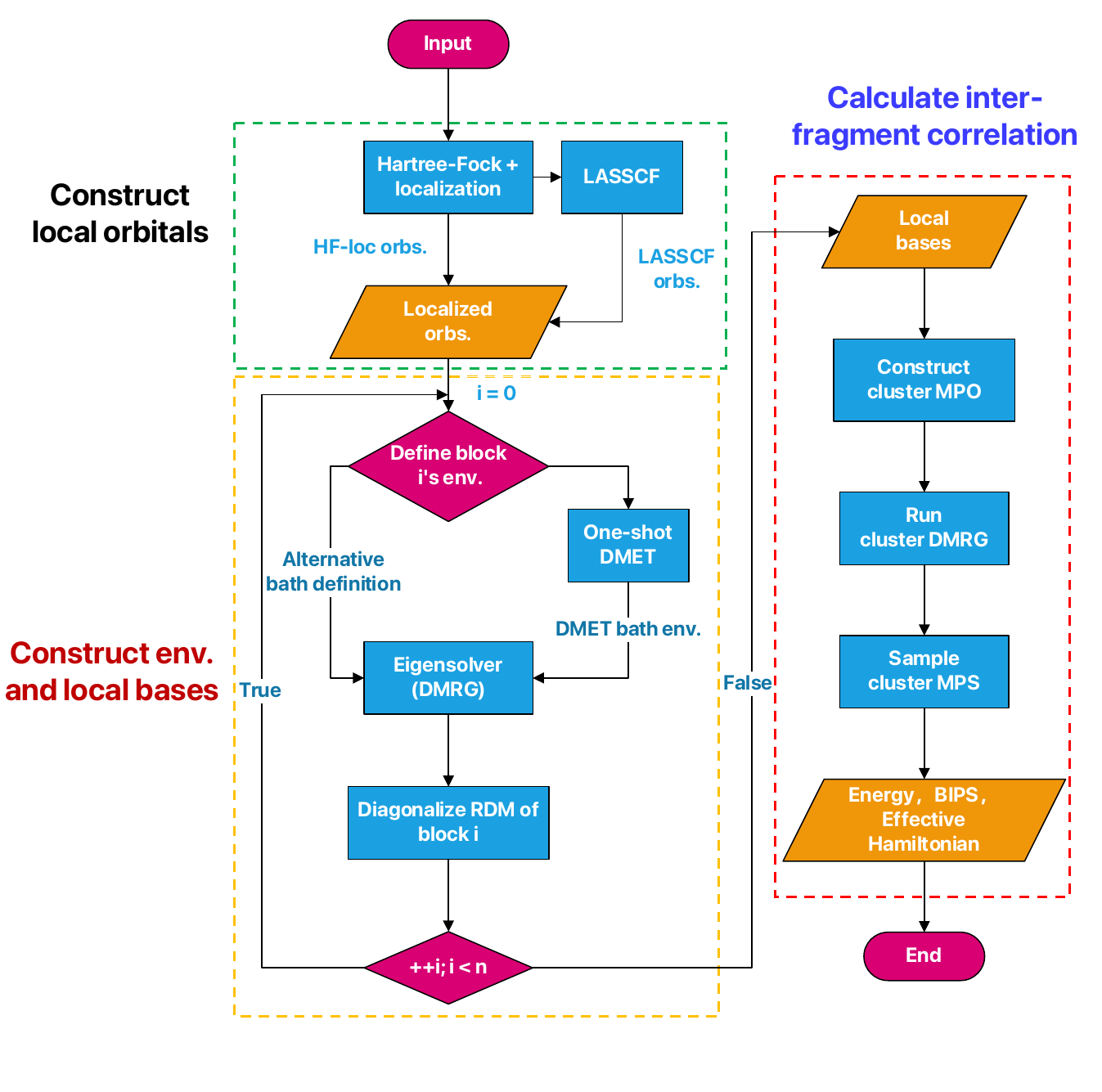}
    \caption{Flowchart illustrating the steps of the BIPS-DMRG framework, where the alternative bath definition refers to other approaches that can generate the bath for the fragment, which can also be incorporated into this framework. For example, using adjacent fragments as bath space.}
    \label{fig:flowchart}
\end{figure}

After obtaining the cMPO, it can be integrated directly into any MPS-based DMRG program to perform cDMRG and obtain the BIPS coefficients \( c_{\alpha_1,\beta_2,\cdots,\omega_N} \) in eq~\eqref{eq:BIPS_form}. The workflow algorithm of the BIPS-DMRG framework is shown in Figure~\ref{fig:flowchart}. Generally, cDMRG operates similarly to the standard DMRG procedure, though some techniques should be considered for efficient implementation. First, the initial guess in cDMRG needs to be addressed. Since the local bases no longer represent only the Hilbert space of single orbitals, the Hartree-Fock configuration or a linear combination of low-energy configurations may not serve as good candidates. Instead, one can use a randomly initialized MPS containing all allowed quantum numbers in each block, or a linear combination of products of "low-energy" local states, which are determined by the diagonal elements of the local Hamiltonian. Another aspect should be noticed is the computational cost of the cDMRG. As noted in previous studies \cite{whiteInitioQuantumChemistry1999,chanMatrixProductOperators2016,daulFullCIQuantumChemistry2000,ma2022density,chan2011density}, the computational cost of the two-site DMRG algorithm is around $d$ times that of the single-site algorithm. For the cDMRG, the difference of computational cost between two-site and one-site will be much larger than that in standard DMRG. Therefore, the one-site DMRG algorithm with perturbative subspace expansion is usually preferred in cDMRG \cite{whiteDensityMatrixRenormalization2005,hubigStrictlySinglesiteDMRG2015}.

Note that spin-adapted cluster-CI procedure (LASSI/TPSCI)\cite{agarawal2024automatic,pandharkar2022localized,abraham2020selected,braunscheidel2023generalization} can also be easily executed using cMPO, since calculating the Hamiltonian
between two BIPSs only requires the tensor contraction between the cMPO and two cluster-MPSs with one auxiliary bond dimension. However, we believe that, after encapsulating strong correlations, the remaining weaker but more homogeneous inter-fragment correlations may be more readily captured by the DMRG procedure. A formal comparison between these two treatments of inter-fragment correlations may require further investigation in the future.


Unlike the comb tensor network (TN), the cMPO in BIPS-DMRG remains a 1D chain rather than a non-1D TN structure. Therefore, it has a much lower computational cost compared to the comb TN. If we ignore the cost of generating the cMPO since it is performed only once, and focus solely on the main contributing factors to the computational cost of standard DMRG, namely \(k\) and \(m\), then the total computational cost of BIPS-DMRG primarily originates from two parts: (1) standard DMRG within the model space \((O(k\tilde{k}^3m^2) + O(k\tilde{k}^2m^3))\) and (2) cDMRG for acquiring the coefficients of the BIPSs \((O(Nk^2 (\tilde{m}^3 + \tilde{m}^2)) + O(Nk^3\tilde{m}^2))\). It is easy to observe the separation between large bond dimensions and a long chain of sites, where the former has high exponents associated with large bond dimensions (\(m^3\)) and the latter has high exponents related to the length of the chain (\(k^3\)). As a result, the scenario in which these two factors collectively lead to high computational costs in DMRG is mitigated by the separation of inhomogeneous electron correlation in BIPS-DMRG. 

However, given the large physical dimensions in cDMRG, its computational cost is also influenced by the dimension of the physical bond, which refers to the number of local states in this context. We present a formal comparison between our BIPS-DMRG and standard DMRG in Table~\ref{tab:computational_cost}. It is evident that the model calculation and cMPO construction will not exceed the time cost of DMRG in the vast majority of cases, whereas the cDMRG procedure may exceed the DMRG calculation when \(n_{state}\) is excessively preserved. Therefore, it should be noted that BIPS-DMRG is not suitable for systems with homogeneous correlation, where \(n_{state}\) is approximately equal to the bond dimension \(m\) used in the DMRG within fragment.

\begin{table*}[htbp]
\resizebox{\textwidth}{!}{
\caption{\label{tab:computational_cost}%
Computational cost of BIPS-DMRG compared to standard DMRG\textsuperscript{\emph{a}}.}
\begin{threeparttable}{
\begin{tabular}{ccccc}
\hline
\hline
\multirow{2}*{Scheme} & \multicolumn{3}{c}{BIPS-DMRG} & \multirow{2}*{standard DMRG} \\
\cline{2-4} & model calculation & construction of cMPO & cDMRG\textsuperscript{\emph{b}} &  \\
\hline
1-site & $O(k\Tilde{k}^3m^2d^2) + O(k\Tilde{k}^2m^3d^2)$ & \multirow{2}*{ \(O((k\tilde{k}^2 + k^2)m^3d) + O(k^4n_{state}^2)\)} & $O(Nk^2 (\tilde{m}^3n_{state} + \tilde{m}^2n_{state}^2)$) + $O(Nk^3\tilde{m}^2n_{state})$ & $O(k^4m^2d^2) + O(k^3m^3d^2)$\\
2-site & $O(k\Tilde{k}^3m^2d^2) + O(k\Tilde{k}^2m^3d^3)$ &  & O($Nk^2 (\tilde{m}^3n_{state}^2 + \tilde{m}^2n_{state}^3)$) + $O(Nk^3\tilde{m}^2n_{state}^2)$ & $O(k^4m^2d^2) + O(k^3m^3d^3)$\\
\hline
\end{tabular}
}
\begin{tablenotes}
\item[a] $k$ is the number of original sites (localized orbitals), \(N\) is the number of clustered sites (fragments), and \(\tilde{k} \coloneqq k/N\) represents the number of sites within a fragment. Additionally, $m$ is the bond dimension used in standard DMRG, $\tilde{m}$ is the bond dimension used in cDMRG, $d$ is the physical dimension of standard DMRG (the dimension of the local Hilbert space for a single orbital), and $n_{state}$ is the physical dimension of cDMRG (the number of local states).
\item[b] The computational cost is derived from this literature \cite{largesite}. To correctly obtain this computational scaling during the implementation, the same technique used in constructing cMPO, which we introduced earlier, must be employed.
\end{tablenotes}
\end{threeparttable}
}
\end{table*}

\subsection{wave function analysis in BIPS-DMRG}
The framework in our work not only enables highly accurate energy calculations but also allows for the definition of several diabatic states (BIPSs). By using an efficient sampling algorithm \cite{leeExternallyCorrectedCCSD2021}, significant BIPSs can be identified, aiding in the analysis of the global wave function's physical meaning. 

Firstly, the final wave function in BIPS-DMRG can be expressed in the right canonical form \cite{schollwaDensitymatrixRenormalizationGroup2011}, which is the most efficient form for sampling though not necessary, as:
\begin{equation}
    \label{eq:cluster_mps}
    |\Psi\rangle = \sum_{\alpha,\beta,\cdots,\omega}\sum_{\{\Tilde{a}_I=1\}}^{\{\Tilde{m}_I\}} {\Tthree[V]{1}{\Tilde{a}_{1}}{\alpha_{1}}\Tthree[V]{\Tilde{a}_{1}}{\Tilde{a}_{2}}{\beta_{2}}\cdots \Tthree[V]{\Tilde{a}_{N-1}}{1}{\omega_{N}}|{\alpha_1\beta_2\cdots\omega_N}}\rangle.
\end{equation}
 The coefficient of any BIPS can then be obtained from the overlap between the BIPS and the state $|\Psi\rangle$ as:
\begin{equation}
    \label{eq:overlap}
    c_{{\alpha_1},\beta_2,\cdots,\omega_N}=
    \langle \alpha_1\beta_2\cdots\omega_N |\Psi\rangle = \sum_{\{\Tilde{a}_I=1\}}^{\{\Tilde{m}_I\}}
    \Tthree[V]{1}{\Tilde{a}_{1}}{\alpha_{1}}\Tthree[V]{\Tilde{a}_{1}}{\Tilde{a}_{2}}{\beta_{2}}\cdots \Tthree[V]{\Tilde{a}_{N-1}}{1}{\omega_{N}}
\end{equation}
However, the computational cost of directly traversing all BIPSs is $O(N_{\text{BIPS}} N \tilde{m}^3)$, where $N_{\text{BIPS}}$ (the number of BIPSs) grows exponentially with the system size. Luckily, during the sweep to compute the coefficient, one can judge the coefficient whether go beyond the threshold at any given site (e.g., at site $P$) by the set of partial coefficients $\{c^{\Tilde{a}_{P}}_{{\alpha_1},\cdots,\gamma_P}\}$, which can be obtained from:
\begin{equation}
\begin{aligned}
    \label{eq:partial_overlap}
    c^{\Tilde{a}_{P}}_{{\alpha_1},\cdots,\gamma_P}=\sum_{\Tilde{a}_{1},\cdots,\Tilde{a}_{P-1}=1}^{\Tilde{m}_{1},\cdots,\Tilde{m}_{P-1}}
    \Tthree[V]{1}{\Tilde{a}_{1}}{\alpha_{1}}\cdots \Tthree[V]{\Tilde{a}_{P-1}}{\Tilde{a}_{P}}{\gamma_P}
\end{aligned}
\end{equation}
Considering the left-normalized condition of left-canonical MPS, the coefficient of BIPS can be written as:
\begin{equation}
\begin{aligned}
    \label{eq:sample}
    c_{{\alpha_1},\beta_2,\cdots,\omega_N}&=\sum_{\Tilde{a}_{P}=1}^{\Tilde{m}_{P}}
    c^{\Tilde{a}_{P}}_{{\alpha_1},\cdots,\gamma_P} \sum_{\Tilde{a}_{P+1},\cdots,\Tilde{a}_{N}=1}^{\Tilde{m}_{P+1},\cdots,\Tilde{m}_{N}}
    \Tthree[V]{\Tilde{a}_{P}}{\Tilde{a}_{P+1}}{\delta_{P+1}}\cdots \Tthree[V]{\Tilde{a}_{N-1}}{1}{\omega_{N}}
    = \sum_{\Tilde{a}_{P}=1}^{\Tilde{m}_{P}}
    c^{\Tilde{a}_{P}}_{{\alpha_1},\cdots,\gamma_P}
    c'^{\Tilde{a}_{P}}_{{\delta_{P+1}},\cdots,\omega_{N}}\\
    &\leq \sqrt{[\sum_{\Tilde{a}_{P}=1}^{\Tilde{m}_{P}}
    (c^{\Tilde{a}_{P}}_{{\alpha_1},\cdots,\gamma_P
    })^2][\sum_{\Tilde{a}_{P}=1}^{\Tilde{m}_{P}}
    (c'^{\Tilde{a}_{P}}_{{\delta_{P+1}},\cdots,\omega_{N} })^2]}\\
    &\leq \sqrt{[\sum_{\Tilde{a}_{P}=1}^{\Tilde{m}_{P}}
    (c^{\Tilde{a}_{P}}_{{\alpha_1},\cdots,\gamma_P
    })^2][\sum_{\delta_{P+1},\cdots,\omega_{N}}
    (c'^{\Tilde{a}_{P}}_{{\delta_{P+1}},\cdots,\omega_{N} })^2]}
    = \sqrt{\sum_{\Tilde{a}_{P}=1}^{\Tilde{m}_{P}}
    (c^{\Tilde{a}_{P}}_{{\alpha_1},\cdots,\gamma_P
    })^2}
\end{aligned}
\end{equation}
following the Cauchy–Schwarz inequality. Therefore, if $\sqrt{\sum_{\Tilde{a}_{P}=1}^{\Tilde{m}_{P}}
    (c^{\Tilde{a}_{P}}_{{\alpha_1},\cdots,\gamma_P
    })^2}$ is smaller than the threshold, then this partial coefficient is dropped, as are all BIPSs involving the local basis string $|\alpha_1\cdots\gamma_P \rangle$.

 Combined with the quantum number information, such as charge and spin quantum number, these important BIPSs can be used to analyze the physical process that global wave function represents. By calculating the expectation of Hamiltonian between these selected BIPSs, the diagonal energies and magnitude of the couplings between different diabatic states can be revealed, enabling the construction of a highly-compressed effective Hamiltonian.
 
\section{Results and discussion}
\begin{figure}[ht]
\includegraphics[width=\textwidth]{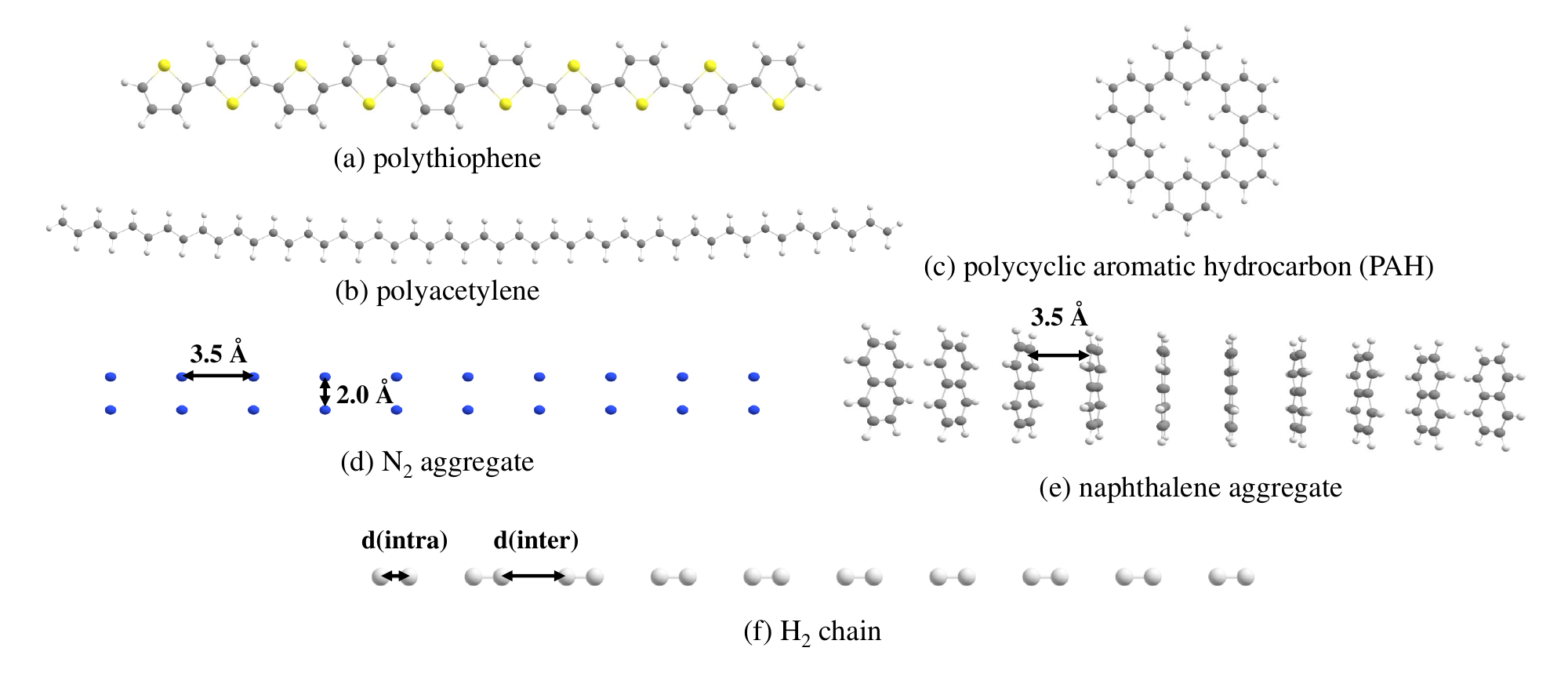}
\caption{\label{fig:fig1} Testbed systems for BIPS-DMRG explored in this work.}
\end{figure}

To validate the BIPS-DMRG method and illustrate the inhomogeneous interactions in chemical systems, we applied it to a range of systems, encompassing both bonding and non-bonding interactions (Figure~\ref{fig:fig1}). The initial localized orbitals for the LASSCF calculations were orthogonal atomic orbitals generated using the meta-L\"{o}wdin method \cite{sun2014exact}. Subsequent calculations, including DMRG, BIPS-DMRG, and ASD-DMRG, were conducted using the optimized LASSCF orbitals. The LASSCF and DMET bath orbitals were prepared with the MRH \cite{hermes2020variational} and PySCF \cite{sun2020recent} packages. All other electronic structure calculations were carried out using the Kylin software package \cite{xie2023kylin}.

\subsection{Accuracy and efficiency tests in ground state}
In this section, we present improvements in accuracy over ASD-DMRG and speed over DMRG. First, we compare the results of BIPS-DMRG with other methods (Table~\ref{tab:table1}), including DMRG, LASSCF, and ASD-DMRG. Here, ASD-DMRG refers to cDMRG utilizing energetically low-lying eigenstates as local states, which are obtained from state-averaged DMRG calculations within each fragment and also represented as MPSs using MPS addition\cite{cheng2024renormalized}. To ensure a fair comparison, we set the parameter \( m \) to 1000 in both global DMRG and the eigensolver DMRG in BIPS-DMRG and ASD-DMRG. Thus, the global DMRG over the whole space serves as the benchmark for the other methods, and local basis states in BIPS-DMRG and ASD-DMRG are represented by the same bond dimension. The threshold for discarded singular values was set to $1 \times 10^{-12}$, ensuring that the preserved states were governed by the bond dimension. Frontier molecular orbitals of the systems were chosen to define the active space, comprising $ \pi $ orbitals for conjugated molecules and 2$p$ atomic orbitals for $\mathrm{N_2}$ molecules. Since our focus was on the active space of these molecules, we primarily employed the minimal basis set STO-3G, except for naphthalene aggregates and polyacetylene, where 6-31G was used, and for $\mathrm{N_2}$ aggregates, where cc-pVDZ was applied.


As illustrated in Table~\ref{tab:table1}, compared to HF, LASSCF significantly reduces the energy deviation and effectively captures the majority of electron correlation (intra-fragment correlation). However, the remaining inter-fragment correlation, which contributes approximately 4-10 $\mathrm{mE_h}$ in non-bonding systems and 30-100 $\mathrm{mE_h}$ in bonding systems, is substantial and cannot be overlooked. It is important to note that in ASD-DMRG, the number of preserved states and their proportions with different quantum numbers must be specified manually. To make a valid comparison, we used information from BIPS-DMRG to define the type and number of preserved states in ASD-DMRG, keeping the number of preserved states across different quantum number sectors identical in both methods. In this situation, we see that ASD-DMRG can indeed improve upon LASSCF by incorporating some energetically low-lying eigenstates; however, it is notably less efficient at recovering inter-fragment correlations compared to BIPS-DMRG, especially in bonding systems. Since in the procedure of cDMRG the physical bond is no longer optimized, the global Hilbert space spanned by products of local states remains the same. The results indicate that BIPS-DMRG identifies more important regions contributing to the ground state in the actual global Hilbert space compared to ASD-DMRG, underscoring the advantage of the BIPS framework, where the preserved local states, accounting for environmental effects, are more compact compared to the isolated local states in ASD-DMRG. Regardless of whether the system is non-bonding or bonding, BIPS-DMRG consistently yields accurate results, with errors ranging from 1-3 $\mathrm{mE_h}$. In addition to the weak inter-fragment correlations resulting in fewer preserved states, these correlations should also lead to a smaller bond dimension in cDMRG. Therefore, these two parameters should be comparable when targeting a single electronic state. For simplicity in this study, we opted to set these parameters as equivalent, i.e., $n_{\text{state}} \equiv \tilde{m}$.

\begin{table*}[htbp]
\resizebox{\textwidth}{!}{
\caption{\label{tab:table1}%
Numerical results of BIPS-DMRG are compared with LASSCF and ASD-DMRG, using the absolute ground state energy of DMRG as the reference. The results of methods other than DMRG are presented in terms of energy errors ($\Delta$) deviating from the reference.}
\begin{threeparttable}{
\begin{tabular}{cccccc}
\hline
\hline
Systems& Reference/$\mathrm{E_h}$\textsuperscript{\emph{a}}& $\Delta$(HF)/$\mathrm{mE_h}$ & $\Delta$(LASSCF)/$\mathrm{mE_h}$ & $\Delta$(ASD-DMRG)/$\mathrm{mE_h}$ & $\Delta$(BIPS-DMRG)/$\mathrm{mE_h}$ \\
\hline
PAH/6-body \textsuperscript{\emph{b}} & -1361.147804 &637.02 & 39.18 & 21.34 & 1.42 \\
polyacetylene/8-body\textsuperscript{\emph{c}} & -1824.751700 & 1021.57& 38.73 & 30.85 &0.35 \\
polythiophene/10-body\textsuperscript{\emph{d}} & -5441.432416 & 771.20& 90.75
 & 13.49  & 0.78 \\
$\mathrm{N_2}$ aggregate/10-body\textsuperscript{\emph{e}} &-1087.896665 & 4595.74 & 4.16
 & 2.12 & 0.99 \\
  naphthalene aggregate/10-body\textsuperscript{\emph{f}} & -3825.754926 & 1743.13 & 7.99  & 7.99 & 2.95 \\
  \hline
\end{tabular}
}
\begin{tablenotes}
\item[a] QC-DMRG with $m$=1000
\item[b] 36 electrons in 36 orbitals, $ n_{\text{state}}$=32, $\Tilde{m}$ = 32
\item[c] 48 electrons in 48 orbitals, $ n_{\text{state}}$=32, $\Tilde{m}$ = 32
\item[d] 60 electrons in 50 orbitals, $ n_{\text{state}}$=64, $\Tilde{m}$ = 64
\item[e] 60 electrons in 60 orbitals, $ n_{\text{state}}$=16, $\Tilde{m}$ = 16
\item[f] 100 electrons in 100 orbitals, $ n_{\text{state}}$=16, $\Tilde{m}$ = 16

\end{tablenotes}
\end{threeparttable}
}
\end{table*}

\begin{figure}[htbp]
\includegraphics[width=\textwidth]{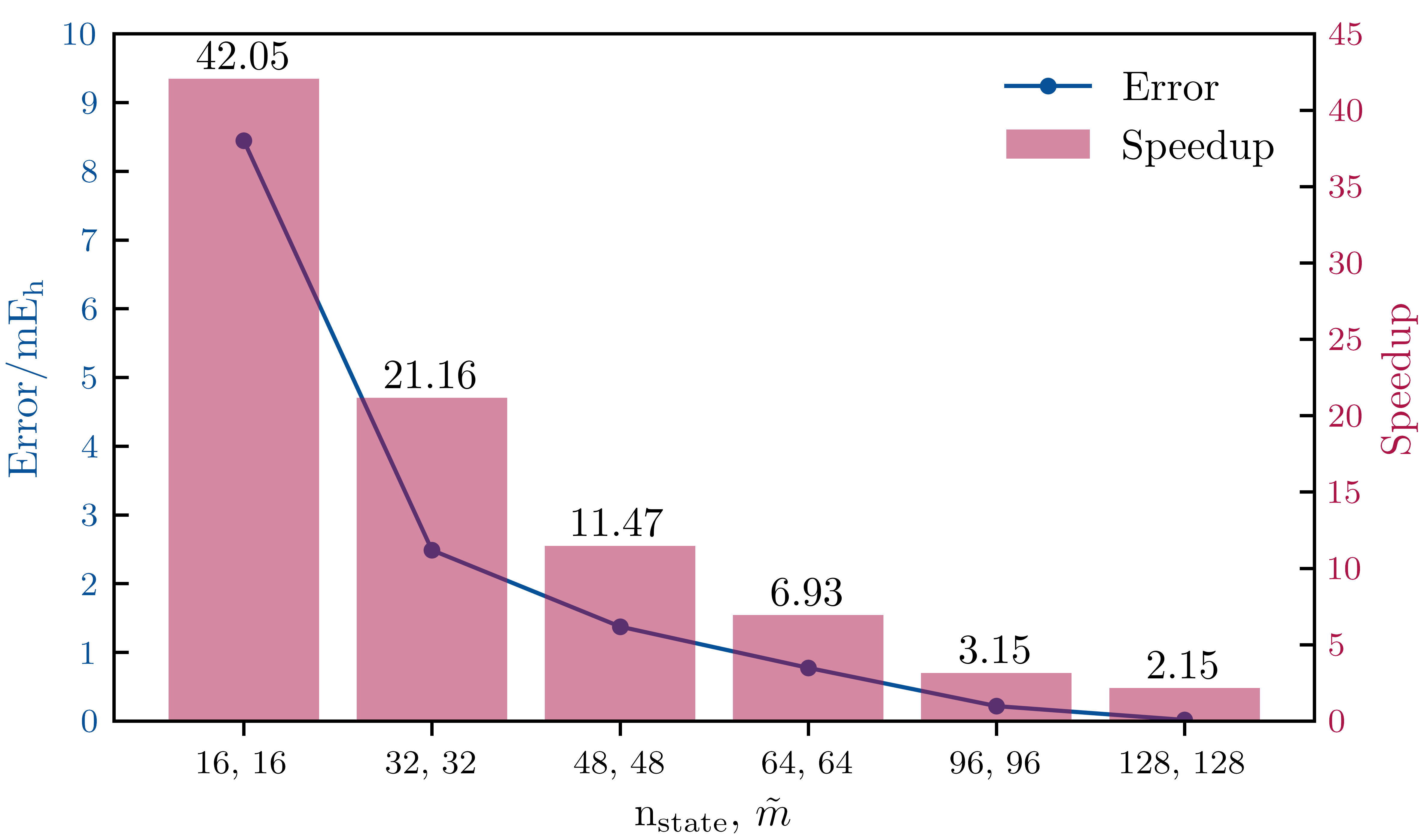}
\caption{\label{fig:speedup}The energy error and speedup of BIPS-DMRG with the number of preserved states ($ n_{\text{state}} $) and the bond dimension ($ \tilde{m} $) for polythiophene.}
\end{figure}

Next, we demonstrated the efficiency of our BIPS-DMRG method compared to global DMRG. We investigated the convergence of BIPS-DMRG with respect to $ \tilde{m} (n_{\text{state}})$ in cDMRG using the polythiophene molecule (Figure~\ref{fig:fig1}.a). The energy error and speedup relative to QC-DMRG are shown in Figure~\ref{fig:speedup}. Speedup was calculated by dividing the wall-clock runtime of QC-DMRG with \( m = 1000 \) by the wall-clock runtime of BIPS-DMRG on the same node, and the absolute time is presented in Table~S1. Here, the computational time of BIPS-DMRG includes the processing procedures, including the one-shot DMET calculation, the DMRG calculation within the model space, and the construction of the cMPO. Additionally, the bond dimension used to represent the local states in BIPS-DMRG is set to 1000. The results indicate a significant reduction in energy error as the number of preserved states increases. Once the number of preserved states reaches 128, the energy error falls below 0.1 mEh, while BIPS-DMRG consistently maintains a speedup of more than 2. 

\begin{figure}[htbp]
\includegraphics[width=\textwidth]{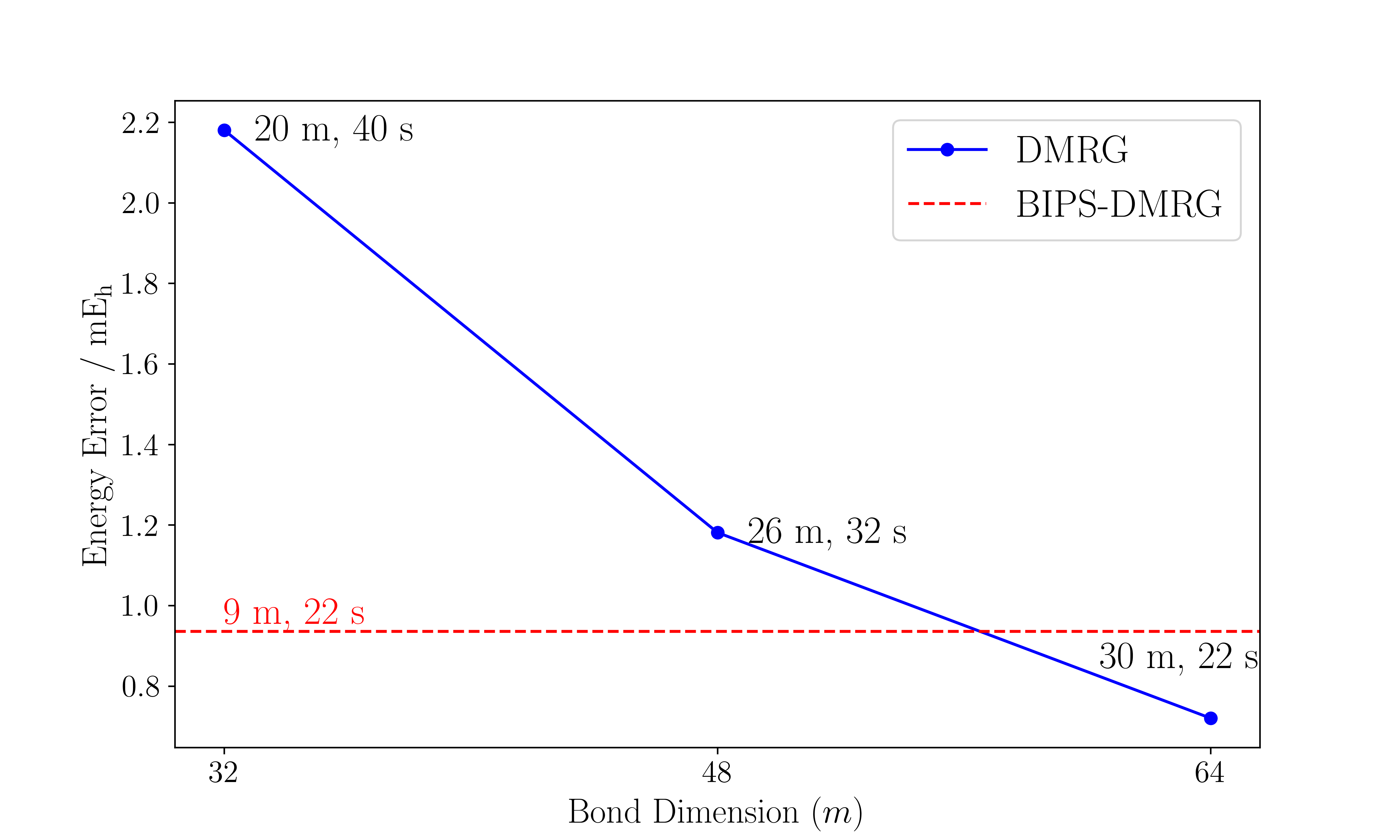}
\caption{\label{fig:n2}The energy error and computational time of BIPS-DMRG and DMRG for N$_2$. The energy error is compared to a reference DMRG calculation with $\tilde{m}=8$ and $m=250$. In DMRG, $\tilde{m}$ represents the bond dimension connecting orbitals associated with different fragments, while $m$ is the bond dimension associated with the other bonds. In all DMRG calculations, $\tilde{m}$ is set to 8, and the x-axis represents the value of $m$. The BIPS-DMRG with $\tilde{m}=8$ and $n_{state}=8$ is shown as a horizontal line.}
\end{figure}

We also compared the BIPS-DMRG with a global DMRG that employs a non-uniform bond dimension. By fixing the bond dimension connecting orbitals associated with different fragments to $\tilde{m}$ and increasing the bond dimension $m$ associated with the other bonds, we investigated the energy error and speedup of BIPS-DMRG compared to global DMRG with such a non-uniform bond dimension over an aggregate of ten N$_2$ molecules (Figure~\ref{fig:fig1}.d). The value of $\tilde{m}$ was set to 8, and the bond dimension used for representing the local states in BIPS-DMRG was set to 250. The results shown in Figure~\ref{fig:n2} indicate that BIPS-DMRG achieves a relatively low energy error in a short time. Although the bond dimension \(m\) increases, allowing DMRG to become more accurate and ultimately outperform BIPS-DMRG, BIPS-DMRG still provides a speedup greater than 2.83x when achieving a similar high accuracy (error within 1 mE\(_h\)).

\subsection{Wave function analysis for multiple states}
Generalizing BIPS-DMRG to simultaneously calculate multiple lowest excited states is straightforward. This is accomplished by diagonalizing the state-averaged reduced density matrix to define renormalized local bases, i.e., using state-averaged DMRG as the eigensolver for the model system. To account for the larger region of the total Hilbert space required to target multiple states, a larger bond dimension in cDMRG is needed. In contrast, the increase in the number of locally preserved states should be minimal or remain the same, as these lowest excited states share many similar local components. Therefore, we kept $n_{\text{state}}$ unchanged and set $\tilde{m}$ to be the number of calculated states multiplied by $n_{\text{state}}$. 

As an illustration,  the results for the first six energetically low-lying states of polythiophene (Figure~\ref{fig:fig1}.a) are shown in Table \ref{tab:table2}.
The energy errors for these six electronic states are still below 10 milli-Hartree, although the excited states exhibit slightly larger errors than the ground state. This is because we used the DMET bath orbitals obtained from the ground state calculation, as defining appropriate bath orbitals for calculations involving multiple states, including excited states, is more challenging than for the ground state. In this study, we did not focus on optimizing the bath orbitals for excited states or applying extrapolation procedures to achieve more accurate results. Instead, we aimed to demonstrate the capability of BIPS-DMRG in handling excited states as a prototype.

\begin{table}[htbp]
\caption{\label{tab:table2}
 State-averaged BIPS-DMRG results for the first six energetically low-lying states of polythiophene, using the energies of state-averaged DMRG as the reference. }
 \begin{threeparttable}
\begin{tabular}{cccc}
\hline
\hline
State	&Reference/$\mathrm{E_h}$\textsuperscript{\emph{a}}	& $\Delta$(BIPS-DMRG)/$\mathrm{mE_h}$\textsuperscript{\emph{b}}\\
\hline
E($\mathrm{S_0}$)&	-5441.432591&	2.32       \\
E($\mathrm{S_1}$)&	-5441.310126&	5.15    \\
E($\mathrm{S_2}$)&	-5441.302995&	4.97   \\
E($\mathrm{S_3}$)&	-5441.294390&	4.83    \\
E($\mathrm{S_4}$)&	-5441.286153&   6.18  \\
E($\mathrm{S_5}$)&	-5441.284758&   6.98   \\
\hline
\end{tabular}
\begin{tablenotes}
    \item[a] State-averaged QC-DMRG with m = 1000
    \item[b] Within the model space, state-averaged QC-DMRG with m = 1000 is used. $ n_{\text{state}}$ = 64 and $\Tilde{m}$ = 6 × 64 = 384 (considering there are six targeted states).
\end{tablenotes}
\end{threeparttable}
\end{table}

By leveraging the clear physical representation of the renormalized local state bases, wave function analysis for the excited states becomes straightforward. We found that the lowest excited electronic states ($\mathrm{S_1}$-$\mathrm{S_5}$) are predominantly doubly excited states, consisting of two triplet excitons, and are therefore referred to as triplet pair states (TT states). It should be noted that dynamic correlations beyond the active space were not considered in the calculation, and as such, the numerical results may differ from realistic observations of this system. However, this does not hinder our analysis of the system within the doubly excited state subspace. Future work will include multi-reference calculations that incorporate dynamic correlations to provide a more comprehensive understanding of the system. Setting a threshold of 0.1 and applying the sample algorithm \cite{leeExternallyCorrectedCCSD2021} introduced in Section~2.4, we found 1, 11, 12, 11, 12, and 10 important BIPSs, respectively, for the six states based on the magnitude of their coefficients exceeding the threshold. In total, 31 important diabatic states were identified, and detailed information can be found in the Supporting Information. 

\begin{figure}
\includegraphics[width=\textwidth]{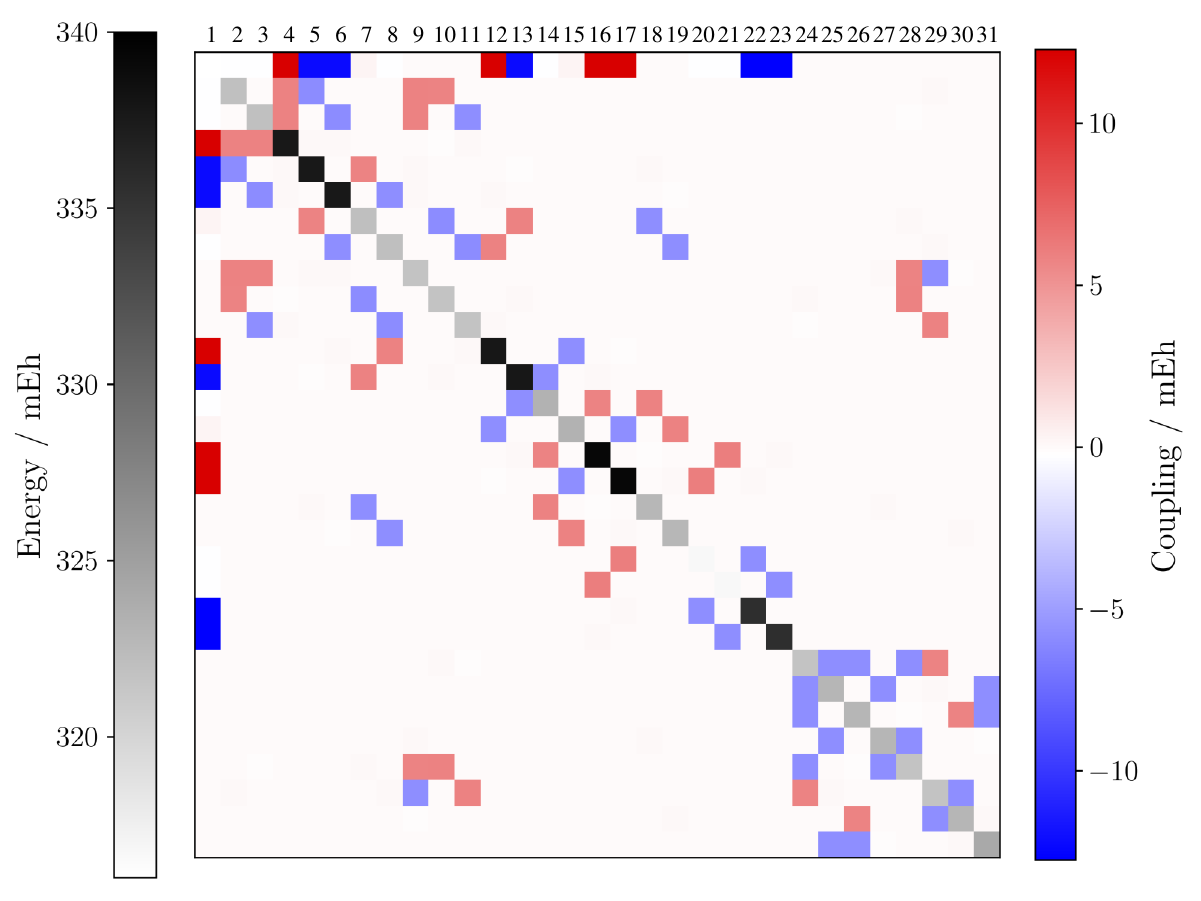}
\caption{\label{fig:effHam} The effective Hamiltonian matrix for polythiophene. The diagonal elements correspond to the excitation energies of diabatic states relative to the $\mathrm{S_0}$ state, ordered as follows: 1.$\mathrm{S_0}$; 2.T(5)T(7); 3.T(4)T(6); 4.T(5)T(6); 5.T(6)T(7); 6.T(4)T(5); 7.T(6)T(8); 8.T(3)T(5); 9.T(4)T(7); 10.T(5)T(8); 11.T(3)T(6); 12.T(3)T(4); 13.T(7)T(8); 14.T(7)T(9); 15.T(2)T(4); 16.T(8)T(9); 17.T(2)T(3); 18.T(6)T(9); 19.T(2)T(5); 20.T(1)T(3); 21.T(8)T(10); 22.T(1)T(2); 23.T(9)T(10); 24.T(3)T(8); 25.T(3)T(9); 26.T(2)T(8); 27.T(4)T(9); 28.T(4)T(8); 29.T(3)T(7); 30.T(2)T(7); 31.T(2)T(9). The numbers in parentheses indicate the corresponding fragment where the triplet state is located. Off-diagonal elements represent couplings between these states.}
\end{figure}

Additionally, we calculated the expectation values of the Hamiltonian elements between these diabatic states by contracting them with the cMPO. The effective Hamiltonian of the multi-excited-state subspace is shown in Figure~\ref{fig:effHam}, which presents a clear physical picture of the electronic structures within the TT manifold, offering valuable insights for research on singlet fission in conjugated systems. It can be observed that the TT states exhibit distinct bifurcation based on their excitation energies. Specifically, adjacent TT states correspond to the darker blocks with higher excitation energies, while long-range TT states are represented by the lighter blocks with lower excitation energies. Among these TT states, only specific TT state pairs that satisfy certain conditions exhibit non-negligible coupling: (1) T(A)T(B) and T(A)T(C) must share the same T exciton located on site A, and (2) sites B and C must be adjacent. For example, the second diabatic state, T(5)T(7), correlates only with the fourth, fifth, ninth, and tenth state--T(5)T(6), T(6)T(7), T(4)T(7), and T(5)T(8), respectively. This coupling can be regarded as triplet excitonic transfer between sites B and C, which closely aligns with Dexter energy transfer. This type of coupling provides a driving force for TT dissociation during singlet fission \cite{smith2010singlet,casanova2018theoretical,miyata2019triplet}.

\subsection{Exploration of various inhomogeneous correlation with $\mathrm{H_2}$ chain}
\begin{figure}
\includegraphics[width=\textwidth]{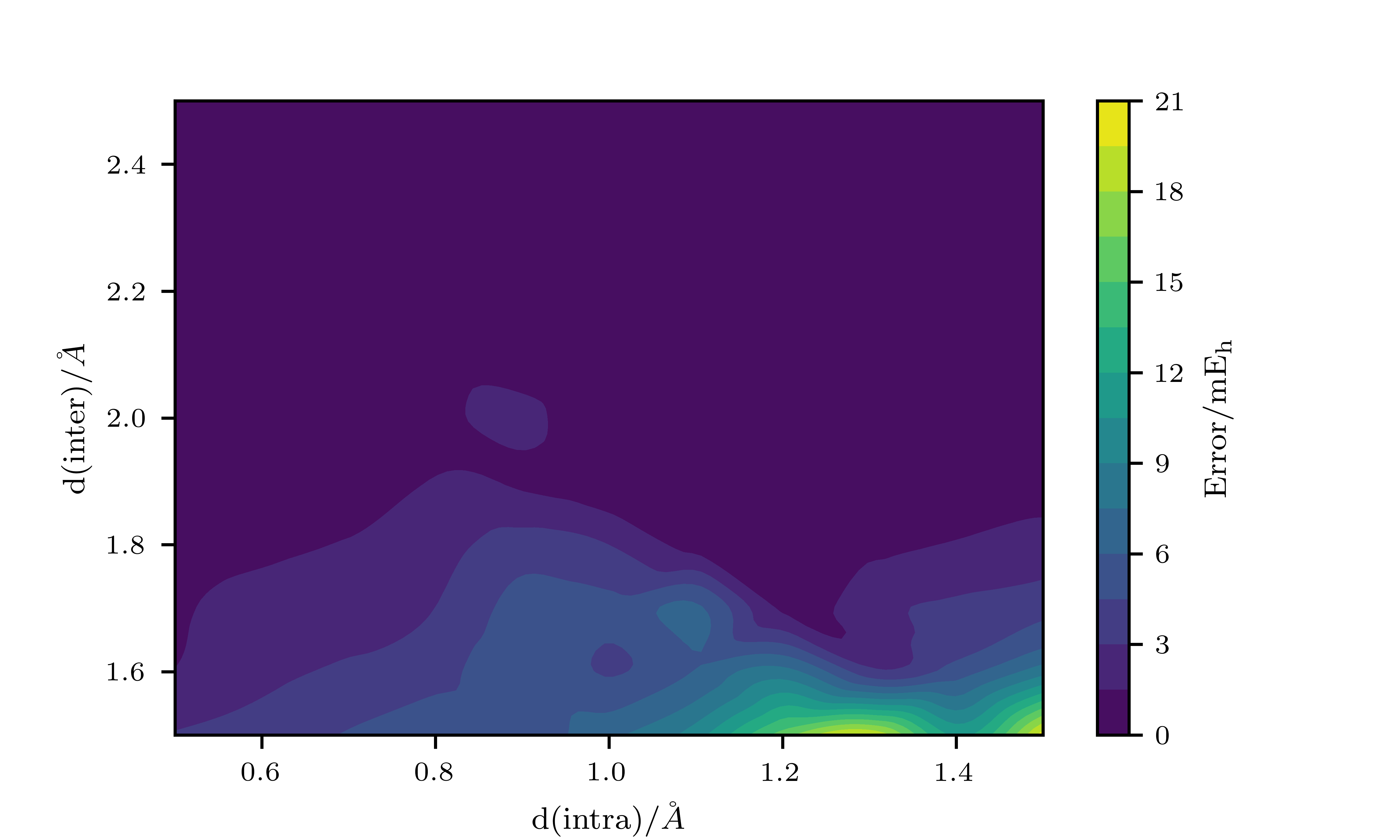}
\caption{\label{fig:h2} Energy error surface for a chain of $\mathrm{H_2}$ molecules relative to standard DMRG. The x-axis and y-axis respectively denote the intra-molecular and inter-molecular distances in the chain (Figure~\ref{fig:fig1}.f).}
\end{figure}

Finally, we assessed the performance of BIPS-DMRG across varying inhomogeneities in a $\mathrm{H_2}$ chain. Figure~\ref{fig:h2} shows the relative energy errors as a function of the intra- and inter-molecular distances in a chain of ten $\mathrm{H_2}$ molecules (Figure~\ref{fig:fig1}.f). Each molecule is chosen as a separate fragment in the analysis. The STO-6G basis set was used, considering the full space of 20 electrons and 20 orbitals. DMRG with a sufficiently large bond dimension ($m = 1000$) was employed as the benchmark. The data were obtained by sampling intra-molecular distances, d(intra), from 0.5 \r{A} to 1.5 \r{A}, and inter-molecular distances, d(inter), from 1.5 \r{A} to 2.5 \r{A}, with a step size of 0.1 \r{A}. Cubic spline interpolation was applied to smooth the two-dimensional data, generating continuous contour lines.

With a maximum of eight preserved states ($n_{\text{state}} = 8$, $\Tilde{m} = 8$), the highest energy error for BIPS-DMRG occurred in the lower right corner of Figure~\ref{fig:h2}, where the intra- and inter-molecular distances are similar, corresponding to the dissociation region of the $\mathrm{H_2}$ molecule. In this scenario, the correlation is relatively uniform, and DMRG is the preferred method, as confirmed by previous studies \cite{chan2011density,hachmann2006multireference,zgid2008density}. In contrast, when the intra-molecular distances are significantly smaller than the inter-molecular distances, the BIPS method efficiently captures the majority of the electron correlation, with errors less than 3 $\mathrm{mE_h}$. The strong performance of BIPS-DMRG in regions with different d(intra) and d(inter), a common occurrence in realistic chemical systems, demonstrates the broad applicability of the BIPS method to molecular systems with inhomogeneities.

\begin{figure}
\includegraphics[width=\textwidth]{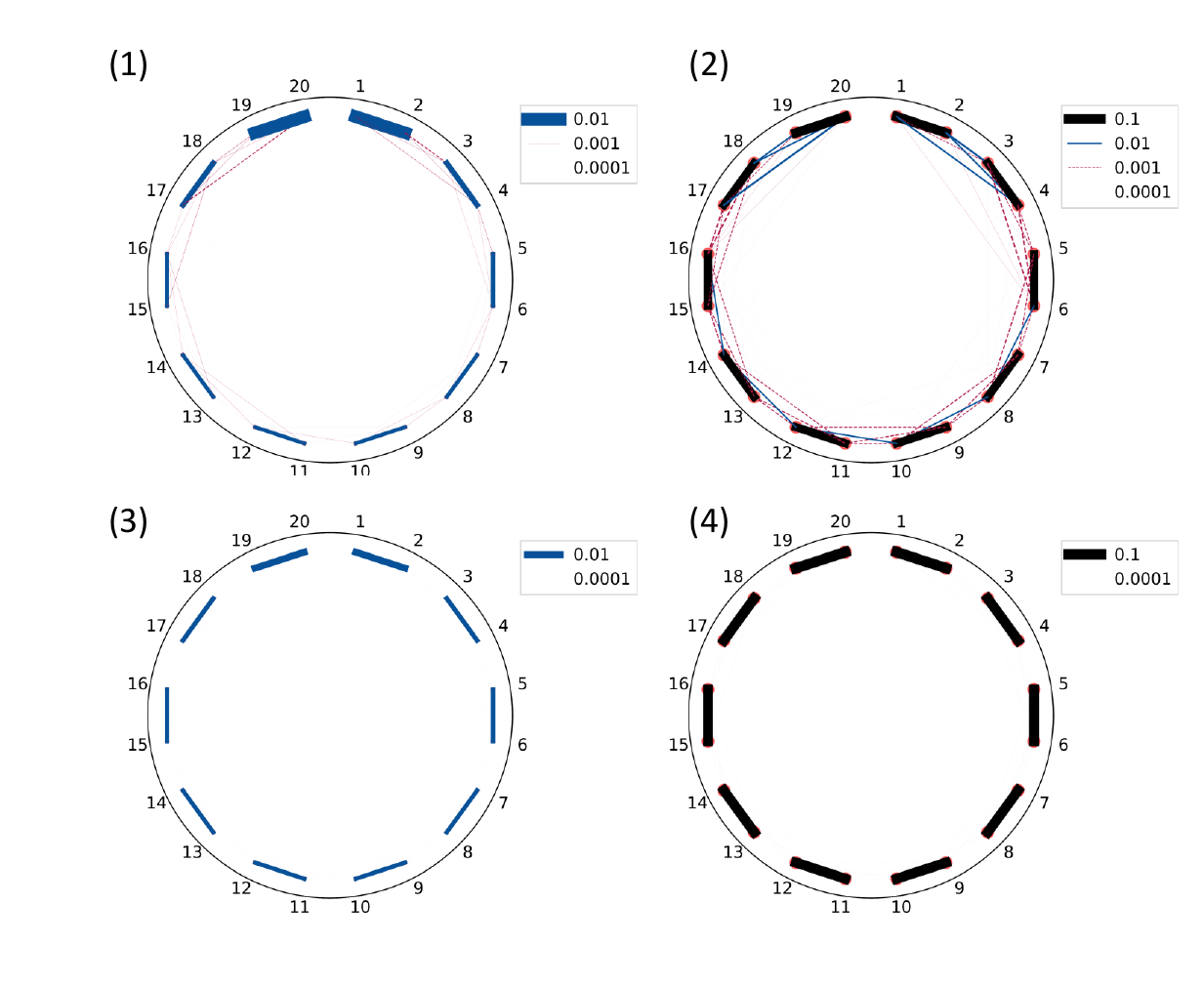}
\caption{\label{fig:MID} Mutual information diagram of the $\mathrm{H_2}$ chain at the four corner regions of Figure~\ref{fig:h2}. Subfigures (1), (2), (3), and (4) correspond to the pairs of distances d(intra) and d(inter): (1) 0.5 \r{A} and 1.5 \r{A}, (2) 1.5 \r{A} and 1.5 \r{A}, (3) 0.5 \r{A} and 2.5 \r{A}, and (4) 1.5 \r{A} and 2.5 \r{A}, respectively. The line widths indicate the strength of entanglement, with widths for entanglement strengths between 0.01-0.1, 0.001-0.01, and 0.0001-0.001 increased fivefold to enhance visibility.}
\end{figure}

We can better understand the performance of BIPS-DMRG by analyzing the mutual information diagram obtained from the QCMaquis package \cite{keller2015efficient}. The four corners of Figure~\ref{fig:h2} were selected to assess the entanglement, with intra- and inter-molecular distances set as: (1) 0.5 \r{A} and 1.5 \r{A}, (2) 1.5 \r{A} and 1.5 \r{A}, (3) 0.5 \r{A} and 2.5 \r{A}, and (4) 1.5 \r{A} and 2.5 \r{A}, respectively. The results, shown in Figure~\ref{fig:MID}, clearly indicate that the inter-fragment entanglement in Subfigures (1) and (2) is significantly larger than in Subfigures (3) and (4), due to the shorter inter-molecular distances. As a result, Subfigures (1) and (2) correspond to the lower left and right corners of Figure~\ref{fig:h2}, where BIPS-DMRG exhibits higher errors.

Several key points are noteworthy. First, in Subfigure (4), the intra-fragment entanglement is as significant as in Subfigure (2), but BIPS-DMRG achieves the same accuracy as in Subfigure (3). This demonstrates that strong intra-fragment entanglements are effectively captured within the BIPS-DMRG framework. Additionally, even in Subfigure (2), where d(intra) equals d(inter), the intra-fragment entanglement remains much larger than the inter-fragment entanglement. This situation contrasts with the entanglement distribution that would be observed using more localized orbitals centered on individual hydrogen atoms rather than hydrogen molecules. This highlights the influence of the localization procedure on entanglement and suggests that further orbital optimization may be beneficial within the BIPS-DMRG framework.

\section{Conclusion}
In summary, we have developed an efficient method called BIPS-DMRG for describing inhomogeneous correlations, which allows us to separately address strong intra-fragment and weak inter-fragment correlations in realistic chemical systems. By employing bath orbitals generated from a one-shot DMET, the environmental effect is considered in the encapsulation of intra-fragment correlation. The remaining weaker, but more homogeneous, inter-fragment interactions can be effectively described by cDMRG using a small bond dimension.

Extensive benchmarking across diverse molecular systems demonstrates that BIPS-DMRG not only surpasses conventional DMRG methods in speed for non-uniform systems, but also achieves superior compression of the local Hilbert space with diabatic states compared to isolated adiabatic eigenstates. This framework can readily accommodate non-Abelian spin-$\mathrm{SU(2)}$ symmetry, making the defined electronic states more physically meaningful. In the future, it will be necessary to develop more refined bath orbitals for multiple open-shell excited states. Consequently, the potential application of BIPS-DMRG to complex assemblies, particularly those involving multiple transition metal and lanthanide/actinide centers, holds great promise for advancing the understanding and study of material properties and catalytic processes.

\begin{acknowledgement}
This work was supported by the National Key R\&D Program of China (2023YFA1506901) and the National Natural Science Foundation of China (Nos. 22325302 and 22073045). The authors thank Prof. Wenjian Liu, Prof. Christian Mendl, Prof. Yang Guo, Dr. Huanchen Zhai,  Dr. Ke Wang and Junjie Yang for stimulating discussions. 

\end{acknowledgement}

\begin{suppinfo}
The following files are available free of charge.
\begin{itemize}
  \item Coordinates of tested chemical systems.
  \item Scripts for generating the LASSCF orbitals and DMET bath orbitals.
  \item Information of diabatic states.
  \item Absolute time of BIPS-DMRG compared to global DMRG for polythiophene.
  \item Mutual information diagram of the $\mathrm{H_2}$ chain with different orbitals.
\end{itemize}

\end{suppinfo}

\bibliography{achemso-demo}

\end{document}